\documentclass[
    ,final            
  ]
  {aipproc}

\layoutstyle{6x9}

\def\lsim{\mathrel{\rlap{\lower4pt\hbox{\hskip1pt$\sim$}}
    \raise1pt\hbox{$<$}}}
\def\gsim{\mathrel{\rlap{\lower4pt\hbox{\hskip1pt$\sim$}}
    \raise1pt\hbox{$>$}}}

\def\bea{\begin{eqnarray}}
\def\bee{\end{eqnarray}}

\begin{document}

\title{Asymptotic behavior of pion form factors}


\author{Mikhail Gorchtein}{
  address={Physics Department, Indiana University, Bloomington, IN 47405}
  ,altaddress={Center of Exploration of Energy and Matter, Indiana
    University, Bloomington, IN 47408}
}

\author{Peng Guo}{
  address={Physics Department, Indiana University, Bloomington, IN 47405}
  ,altaddress={Center of Exploration of Energy and Matter, Indiana
    University, Bloomington, IN 47408}
}

\author{Adam P. Szczepaniak}{
  address={Physics Department, Indiana University, Bloomington, IN 47405}
  ,altaddress={Center of Exploration of Energy and Matter, Indiana
    University, Bloomington, IN 47408}
}

\begin{abstract}
 We consider the electromagnetic and transition pion form factors. 
Using dispersion relations we simultaneously describe both the
hadronic, 
time-like region and the asymptotic region of large energy-momentum transfer. 
For the latter we propose a novel mechanism  of Regge fermion
exchange. 
\end{abstract}

\maketitle


Photons interact with quarks,  the charged constituents of hadrons and the resulting electromagnetic form factors probe the quark energy-momentum distribution in hadrons. 
 In this article we  examine  the charged pion electromagnetic form
 factor $F_{2\pi}(s)$, which is   defined  by the matrix element  
$\langle \pi^+(p') \pi^-(p) |J_\mu |0 \rangle = e(p' -p)_\mu F_{2\pi}(s)$,  
 and the transition from factor 
between the neutral pion and a real photon, $F_{\pi\gamma}(s)$ determined  by $\langle \pi^0(p')\gamma(\lambda, p) |J_\mu |0 \rangle = 
 ie^2/4\pi^2 f_\pi  \epsilon_{\mu \nu \alpha\beta} \epsilon^{*\nu}(\lambda) p'^\alpha p^\beta  F_{\pi\gamma}(s) $. 
 Above, $J_\mu$ is the electromagnetic current,  $s = (p' + p)^2$ is
 the four-momentum transfer squared and $f_\pi = 92.4\mbox{ MeV}$ 
is the pion decay constant. Current conservation implies 
 $F_{2\pi}(0) = 1$ and, in the chiral limit, axial
 anomaly determination of the  $\pi^0 \to 2\gamma$ decay leads to the
 expectation $F_{\pi\gamma}(0) \approx 1$.  
Because  at short distances quark/gluon interactions are
asymptotically free, 
it has been postulated  that at high energy or momentum transfer,
both form factors measure 
   hard scattering  of the photon with a small number of the QCD constituents~\cite{EfremovRadyushkin1980,DuncanMueller1979,LepageBrodsky1979}.
The available data
on the pion electromagnetic form factor ranges up to 
$|s|  \lsim 10 \mbox{ GeV}$ \cite{Aubert2009a}
and is approximately a factor of three above 
the asymptotic prediction \cite{BrodskyFarrar1973}.  
  Even more spectacular discrepancy is observed  in  the  transition form factor
  recently measured by  BaBar~\cite{Aubert2009b}.  
 For momentum transfers 
as  large as  $-s \approx 40 \mbox{GeV}^2$ 
the data suggest that the magnitude
  of  $-s F_{\pi\gamma}(s) $ grows with $|s|$, 
whereas pQCD
  predicts $ s F_{\pi\gamma}(s) \to 2 f_\pi$  as $|s| \to
  \infty$ \cite{LepageBrodsky1979}.
 
In the following, we relate the form factors in the 
space-like ($s < 0$) and time-like ($s > 0$) regions
through a dispersion relation (DR), 
 and focus  on the  dynamics in the asymptotic region, $s
\to + \infty$. 
In view of the BaBar "anomaly"  and the apparent failure of the pQCD 
description, we propose a novel description 
 for the dominant mechanism that drives the asymptotic behavior of the
 form factors \cite{Gorchtein2011a}. 

The discontinuity   of
  $F_{\pi\gamma}(s)$ across the unitary  cut  is given by 
\begin{equation} 
{\rm Im}\, F_{\pi\gamma} = t^*_{2\pi,\pi\gamma} \rho_{2\pi} F_{2\pi} +  t^*_{3\pi,\pi\gamma} \rho_{3\pi}  F_{3\pi} + 
\sum_{X\neq2\pi,3\pi} t^*_{X,\pi\gamma} \rho_X F_X .  \label{imt}
\end{equation} 
\indent
 Here,  $t_{X,\pi\gamma}$ ($F_{X}$) represent the amplitudes for 
$X\to \pi^0 \gamma$ ($\gamma^* \to X$), respectively and 
$\rho_{X}$ is a product of the phase space and kinematical factors 
  ({\it i.e.}  for the $2\pi$ intermediate state  $\rho_{2\pi}(s) = s (1- s_{th}/s)^{3/2}/96\pi$).
 Provided Im$\,F_{\pi\gamma}$ vanishes at $s\to\infty$, its real
 part can be reconstructed for any $s$ from the unsubtracted DR 
  \begin{equation} 
 F_{\pi\gamma}(s) = \frac{1}{\pi} \int_{s_{th}}  ds' \frac{Im
   F_{\pi\gamma}(s')}{s' - s}. 
\label{disp}
 \end{equation} 
The two lowest mass intermediate states, $X=2\pi,\,3\pi$ that are dominated 
  by the $\rho(770)$ and $\omega(782)$ resonances, respectively, to a
  large extent 
saturate the cut in the hadronic range $s_{th} < s  \lsim  1 \mbox{GeV}^2$.  
The contribution of a narrow  resonance 
to $F_{\pi\gamma}$ can be well approximated by a 
Breit-Wigner distribution,
 \begin{equation} 
 F_{\pi\gamma}^{V}(s)  =  
c^{V}_{\pi\gamma} m_V^2/
[m_V^2 - s - i m_V \Gamma_V(s)].
\label{omega} 
\end{equation}  
\indent
We obtain $c^{(3\pi)}_{\pi\gamma}   = c^{\omega}_{\pi\gamma}   = 4\pi^2 f_\pi g_{\omega\pi\gamma}/ m_\omega g_\omega = 
      0.493 $ (obtained with $\omega \to \pi\gamma$ 
and $\omega \to e^+e^-$ decay widths yielding 
       $g_{\omega\pi\gamma} = 1.81$ and $g_\omega = 17.1$,
       respectively), and 
$c^{(2\pi)}_{\pi\gamma}   = c^{\rho}_{\pi\gamma}   = 4\pi^2 f_\pi g_{\rho\pi\gamma}/ m_\rho g_\rho = 
      0.613 $ 
(with $\rho \to \pi\gamma$ and $\rho \to e^+e^-$ decay widths leading to  
       $g_{\rho\pi\gamma} = 0.647$ and $g_\rho = 4.96$). 
The isovector contribution can be further improved 
   using a unitary parametrization of  \cite{PhamTruong1976}.
At higher energies, $s \gsim 1{ GeV}^2$ the $K{\bar K}$ 
inelastic channel 
and other multi-particle intermediate states are expected to contribute. 
Unfortunately, since no time-like data are available one cannot 
unambiguously determine these contributions. 
A possible determination of the multi-particle hadronic  states could 
 be given in terms of quark/gluon intermediate states.

\begin{figure}[h]
\centering
\vspace{-0.5cm}
\includegraphics[width=14cm]{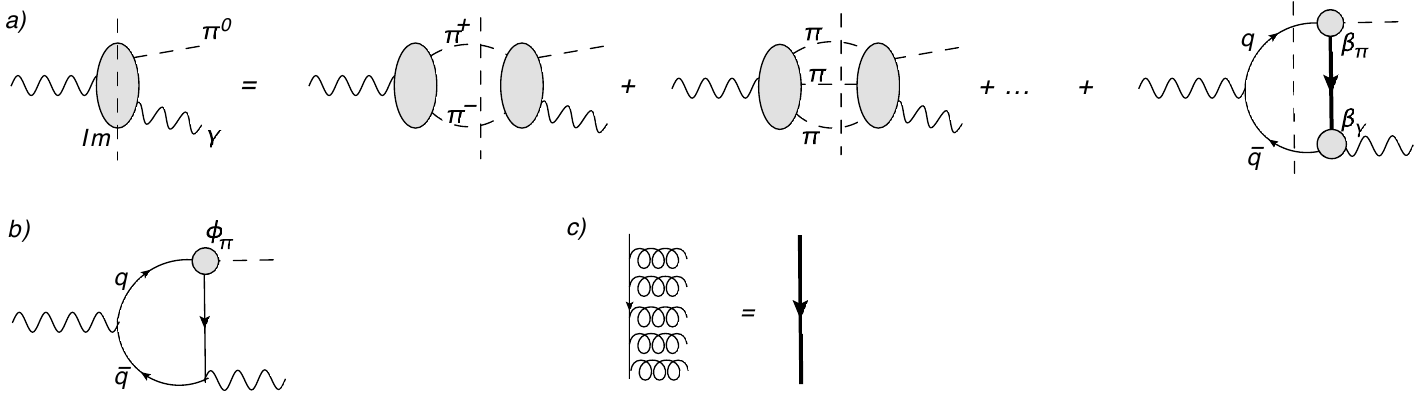}
\caption{Hadronic and asymptotic contributions to the $\pi^0$
  transition form factor.}
\label{fig1}
\end{figure}

\begin{figure}[h]
\vspace{-0.5cm}
\includegraphics[width=14cm]{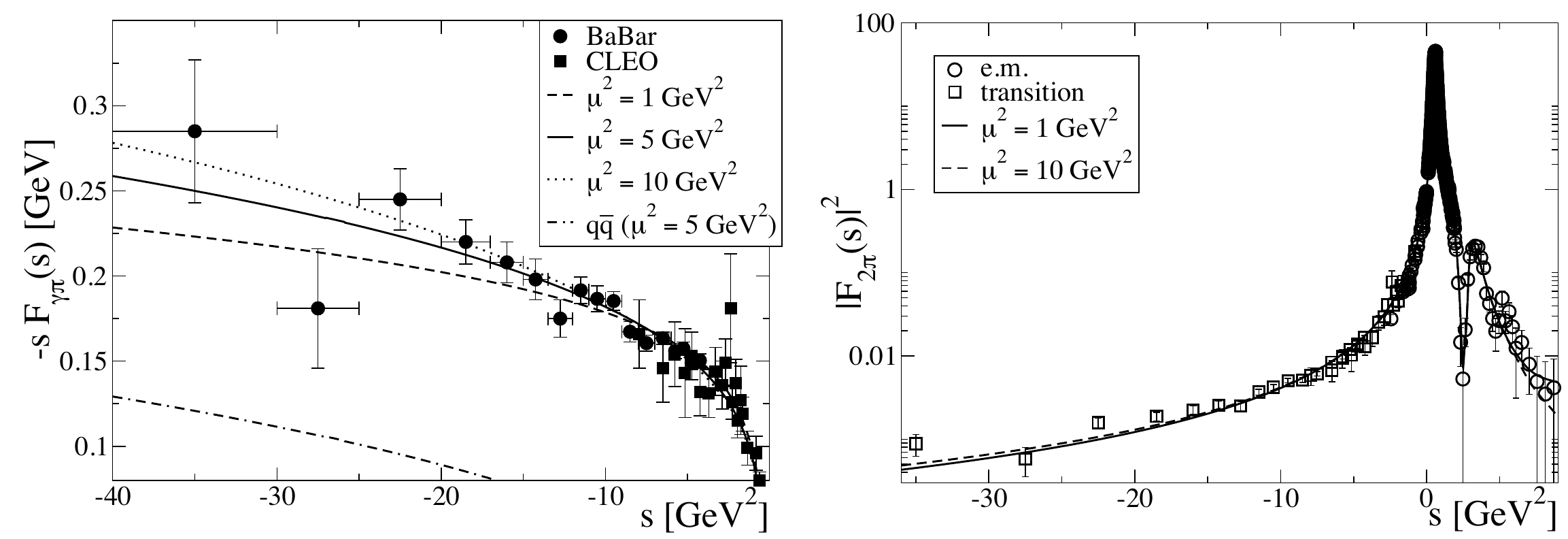}
\caption{Left panel: our results for $|F_{\pi\gamma}(s)|$ in the
  space-like region for  $\mu^2=$ 1 GeV$^2$ (dashed line), 
5 GeV$^2$ (solid line), 10 GeV$^2$ (dotted line),  in comparison with
the experimental data from \cite{Aubert2009b,Gronberg1998}. The Regge contribution
with $\mu^2=$ 5 GeV$^2$ is shown separately (dash-dotted line).
Right panel: our results for the pion electromagnetic form factor for $\mu^2=1$ GeV$^2$
  (solid line) and $\mu^2=10$ GeV$^2$ (dashed line) vs. experimental
  data on the time-like and space-like e.-m. form factor from \cite{Aubert2009a}
  (solid circles).} 
\label{fitt}
\end{figure}

Since the electromagnetic 
form factor $F_X$ of a composite state decreases with energy-momentum transfer,    
asymptotically the {\it r.h.s}  of Eq.~(\ref{imt}) is dominated 
by the $X = q \bar q$, quark-antiquark intermediate state. Its 
form factor is $F_{q\bar q}=1$, and the state contributes to 
Im$\,F_{\pi \gamma}$ via the $q \bar q \to \pi \gamma$, 
$P$-wave  scattering amplitude, $t_{q\bar q,\pi\gamma}$ as illustrated  
by the last diagram in Fig.\ref{fig1}a.  The $q\bar q$ contribution 
shown in  Fig.\ref{fig1}a may be compared to the one in  Fig.\ref{fig1}b,  which 
represents the  asymptotic contribution as predicted by pQCD.   In the 
latter, the $q\bar q \to \pi\gamma$ scattering amplitude, shown to the
right of the vertical cut line, is given by a free quark propagator 
exchanged between the final state pion and photon.  In 
the kinematically relevant  domain  $s\gg t $, $t$
 being the momentum squared carried by the exchanged quark,  
the amplitude $t_{q\bar q,\pi\gamma}$ is expected to have a Regge
behavior \cite{FadinSherman1977}
$ t_{q\bar q,\pi\gamma}(s,t)  = \beta_\pi(t) \beta_\gamma(t) s^{\alpha_q(t)} \approx 
  e^{ b t } s^{\alpha_q} $, where $\beta_\pi,\beta_\gamma$ are
  residues of the exchanged quark at the corresponding vertex.
The difference between the free, Fig.\ref{fig1}b and the Regge propagator  
Fig.\ref{fig1}a can originate 
from the sum of ladder gluons in the wee region ({\it cf.} Fig.\ref{fig1}c).   
The quark Regge trajectory $\alpha_q(t)\approx\alpha_q(0)+\alpha'_qt$ 
is not known; however, 
phenomenologically it can be related to the leading Regge exchange  in $\pi\pi$ scattering.
Identifying reggeized $\rho$ and $f_2$-exchanges with reggeized $q\bar
q$-exchanges then leads to
the expectation $A_{q\bar q}\sim s^{2\alpha_q(t)-1}$, or 
$\alpha_q(t) \sim  \frac{1}{2} (\alpha_\rho(t) + 1)   \approx 0.75+0.45 t/GeV^{2}$.
Further  analysis of  phenomenological implications of 
this quark reggeization will be given in the forthcoming paper~\cite{Gorchtein2011}.
After projecting onto the $p$-wave, the energy dependence 
of the asymptotic, $q\bar q$ contribution to $Im F_{\gamma\pi}$ is 
therefore expected to behave as
   \begin{equation} 
   {\rm Im}\,F^{(q\bar q)}_{\gamma^*\pi\gamma}(s) \to  c^{(q\bar q)}_{2\pi}  (s/GeV^2)^{\alpha_q(0) - 3/2}. 
       \label{ass} 
     \end{equation}  
\indent
Combining the $\omega$ 
and the $\rho$ resonance contributions of Eq.~(\ref{omega})  with 
the asymptotic  form of Eq.~(\ref{ass}) and making the  simplifying 
assumption that the first two contribute to $Im F_{\pi\gamma}$ 
for $s< 1\mbox{ GeV}^2$ while the asymptotic part saturates 
$Im F_{\pi\gamma}$ for $s > \mu^2 $, we fit the available data 
using Eq.~(\ref{disp}) with the single free parameter $c^{(q\bar q)}_{\pi\gamma}$ 
 that  determines the normalization of the asymptotic contribution.  
The result is shown in the left panel of Fig.\ref{fitt}. It is worth noting that even
at largest values of $-s$ the bare $q\bar q$ production gives only
about $50\%$ (dash-dotted line in the left panel of Fig. \ref{fitt}) of the form factor with the remaining half 
coming from the resonances.

In the case of the pion electromagnetic form factor, the discontinuity reads 
\begin{equation} 
{\rm Im}\, F_{2\pi} = t^*_{2\pi,2\pi} \rho_{2\pi} F_{2\pi} +  t^*_{K\bar K,2\pi} \rho_{2K}  F_{K} + 
\sum_{X} t^*_{X,2\pi} \rho_X F_X  
\label{uem}
\end{equation} 
where in the sum is over 
 intermediate states ($X\ne 2\pi, K\bar K$)  in 
$\gamma^* \to X \to 2\pi$ and where we separated 
 the two channels  $X=2\pi$ and $X=K\bar K$  
which phenomenologically are most significant in the hadronic 
domain.
Above the inelastic threshold,  $s> s_{i}$, the unitarity relation now 
involves both  $Im F_{2\pi}$ and $Re F_{2\pi}$ and can be solved 
algebraically. Assuming that the elastic amplitude, $t_{2\pi,2\pi}$ 
asymptotically  approaches the diffractive limit, 
$t_{2\pi,2\pi} \to i/2\rho_{2\pi}$, from Eq.~(\ref{uem}) one finds 
 \begin{equation} 
 F_{2\pi}(s) \to 2 i \sum_{X \ne 2\pi} t_{X,2\pi} \rho_X F^*_X \to 2 i t_{q\bar q,2\pi} \propto i s^{\alpha_q(0)-3/2}. 
 \end{equation} 
\indent
Except for the expected energy dependence, 
we do not know separately the real and  imaginary parts of $t_{q\bar q,2\pi}$. Assuming, as in the case 
of the transition form factor, that the real part of the 
discontinuity due to $q\bar q$  state has the energy dependence 
given by the reggized quark exchange, we can compute $F_{2\pi}$ using
Eq.~(\ref{uem}) and the Cauchy representation.
We approximate the sum over inelastic channels by the single 
 $K{\bar K}$ channel,  and  above $s \ge \mu^2$ the residual $q\bar q$ continuum with 
\begin{equation} 
 {\rm Re}\, t^*_{q\bar q,2\pi} \rho_X = c^{(q\bar q)}_{2\pi}
 (s/GeV^2)^{\alpha_q(0)-3/2}. 
\label{asem} 
\end{equation} 
\indent
For the $t_{2\pi,2\pi}$ and $t_{K\bar K,2\pi}$ amplitudes we use the parametrization
 from ~\cite{Guo2010}. 
We parametrize the isovector kaon form factor $F_{K}$ using 
Breit-Wigner distributions  which include the  $\rho(770)$, 
$\rho'(1400)$ and  $\rho''(1700)$ \cite{FelicettiSrivastava1981}.  Finally we fit the available 
data on $|F_{2\pi}(s)|^2$ with five parameters: the magnitude and 
phase of the $\rho'$ and $\rho''$ contributions to $F_{K}$ 
and $c^{(q\bar q)}_{2\pi}$ --the magnitude of the $q\bar q$ continuum, 
Eq.~(\ref{asem}). 
 In the right panel of Fig.~\ref{fitt}, we display our results for the electromagnetic pion form 
factor $F_{\pi}$ in the range -40 GeV$^2\leq s\leq10$ GeV$^2$. We
confront them with the available experimental data for the
electromagnetic form factor for -10 GeV$^2\leq s\leq10$ GeV$^2$ and
the transition form factor for -40 GeV$^2\leq s\leq-0.8$ GeV$^2$
(both are normalized to 1 at $s=0$). First, we note that in the
space-like region the data sets for the two form factors look
identical. 
One can see that our model describes all the
available data throughout the shown kinematics. 
In the case of the electromagnetic form
factor, our result is a prediction for the $s$-dependence at large
$|s|$, where no data exist so far. In particular, we predict that, as 
for the transition form factor case, $|s\,F_{2\pi}(s)|$ has to rise
asymptotically roughly as $s^{1/4}$, unlike pQCD predictions that
feature at most a logarithmic limit for that combination. 

 This work was supported in part by the US
Department of Energy under contract DE-FG0287ER40365 and the
National Science Foundation grants PHY-0555232 (M.G.), PIF-0653405 (P.G.).

\def\etal{\textit{et al.}}

\end{document}